\documentclass[a4paper]{article}
\usepackage[utf8]{inputenc}
\usepackage{hyperref}

\title{Reconsidering the horizon problem}
\author{Lucas Lolli Savi \\ \href{mailto:lou@uel.br}{lou@uel.br} \\ Departamento de Física, Universidade Estadual de Londrina \\ Campus Universitário, Londrina, PR, Brazil, 86057-970}

\begin{document}

\maketitle

\begin{abstract}
	It is argued that the spatial homogeneity of the early universe need not be explained as having evolved from a previous inhomogeneous state. Rather, a homogeneous cosmic initial condition
	follows from the Second Law of Thermodynamics applied to gravitating systems.
\end{abstract}

Ever since the pioneering 1965 antenna detection by Penzias and Wilson\cite{pw} and all the the way up to the recent Planck mission\cite{planck}, it has been known, ever more certainly and precisely, that the cosmic microwave background (CMB) is highly isotropic as seen from our location in the cosmos. Since there is no good reason why such location should be privileged in any sense, for those five decades mainstream cosmology has been interpreting this fact as indicating that the early cosmic plasma was a very homogeneous fluid.

In 1979, Dick and Peebles\cite{dp}, the same theoreticians who had predicted Penzias and Wilson's detection, argued that such homogeneity was an enigma, given that distant regions of the last scattering surface had not had enough time to communicate. They did not however propose any idea to explain the fact.

That puzzle was addressed by Guth\cite{guth}, who called it the \emph{horizon problem}, in a famous article published two years later. Therein, a new theory called \emph{cosmic inflation} was presented which rests on the key assumption that in the very early universe there was a mechanism which triggered an exponential expansion, {\it i.e.}, inflation. That would mean that the portion of space from which the presently detected CMB was emitted would correspond to such a small region before the onset of inflation that all its parts would have had time to communicate even in the short time window before inflation started. Such communication would have led to the mentioned homogeneous state by means of thermalization, solving the horizon problem.

The above story summarizes the mainstream view on the matter of the isotropic CMB. Contrary to that view, I intend to argue in this paper that the isotropic CMB need not be seen as a puzzle. Quite in the opposite direction, it is in fact a predictable consequence of the theory of thermodynamics of gravitating systems, and constitutes evidence in its favor.

The first thing to note is that, in the light of modern Big Bang cosmology, the old question ``why are there irreversible processes?" was reduced to ``why is the cosmic boundary condition at the Big Bang one of low entropy?". That understanding was a great advance in the history and philosophy of physics\cite{price}. To pursue the matter any further, however, one must know what constitutes a low entropy cosmic configuration.

In the systems most commonly studied in elementary thermodynamics --- gases devoid of long-range forces --- collisions are the only interactions at play, giving rise to pressure in the macroscopic domain. Since collisions are repulsive, such a system tends to occupy all available space and distribute as evenly as possible. Thermal equilibrium --- the state of maximum entropy --- means in this case a homogeneous configuration.

That is not so, however, when other interactions come to play. Gravity, in particular, has an attractive effect (as far as matter density dominates over the cosmological constant, as is currently the case in astronomical scales, and as was the case at the time of last scattering in the scale corresponding to the present observable universe). Thus, while a gas of molecules in a box tends to spread to fill the box they are in, a gas of stars in the corresponding situation would tend to contract down to a point.

In old Newtonian gravity, one could argue that the above parallel would not have been fair, since, disregarding the internal structure of stars and treating them as point masses, they could in principle bounce elastically and go back the way they came. The process would be a reversible one, no entropy being created. However, in general relativity, the collapse of such a system would form a black hole, an irreversible process with a huge entropy increase from the point of view of a stationary observer due to the hiding of information behind an event horizon. And this is exactly the crucial point: event horizons have entropy, and the gravitational clumping of matter creates them. That is known since Bekenstein's\cite{bek} and Hawking's\cite{hawk} studies of black hole thermodynamics in the 1970's. Quantitatively, the entropy of an event horizon is given by\cite{hawk}
\[
S = \frac{1}{4} k \frac{A}{l_P^2},
\]
where $k$ is Boltzmann's constant, $A$ is the area of the horizon and $l_P$ is the Planck length:
\[
l_P = \sqrt\frac{\hbar G}{c^2}.
\]
The fact that event horizons have entropy is what makes the parallel with the gas in a box justified: while non-gravitating systems increase their entropy by going from inhomogeneous configurations to homogeneous ones, gravitating systems, on the other hand, increase their entropy by going from \emph{homogeneous} configurations to \emph{inhomogeneous} ones. That is precisely how structure is formed in our universe.

Since we know that entropy is ever increasing and that homogeneous states correspond to low entropy in gravitating systems, we must conclude from the above that the universe started off very homogeneous. So the isotropy of the CMB should not come as a surprise: it is precisely what one should \emph{expect} to find when looking at early cosmic times, in light of our current knowledge of gravity and thermodynamics.

There is surely still meaning in asking what determines the cosmic boundary condition, and why it must be a homogeneous one. That is a tricky question, if anything because the very existence of a cosmic time parameter, including a Big Bang on which to impose boundary conditions, is itself tied to a maximally symmetric spatial foliation, pointing to a possible consistency loop. In any case, the specific idea that prior thermal contact must have been necessary to achieve homogeneity across the observable universe at the time of last scattering deserves to be reconsidered.

\section*{Acknowledgment}
I would like to thank Guilherme Franzmann for useful comments on an early draft.


\begin{thebibliography}{999}
	\addcontentsline{toc}{chapter}{References}

\bibitem{pw}
A. A. Penzias \& R. W. Wilson,
{\it A Measurement of Excess Antenna Temperature at 4080 Mc/s},
ApJ {\bf 142}, 419--421 (1965)
\url{http://dx.doi.org/10.1086/148307}

\bibitem{planck}
P. A. R. Ade et al.,
{\it  Planck 2015 results, XIII. Cosmological parameters},
A\&A {\bf 594}, A13 (2016)
\url{https://doi.org/10.1051/0004-6361/201525830}

\bibitem{dp}
R. H. Dicke \& P. J. E. Peebles,
{\it The big bang cosmology: enigmas and nostrums},
in S. W. Hawking \& W. Israel, eds.,
{\it General Relativity: An Einstein Centenary Survey}
(Cambridge University Press, London, 1979)

\bibitem{guth}
Alan Guth,
{\it Inflationary universe: A possible solution to the horizon and flatness problems},
Phys. Rev. D {\bf 23}, 347--356 (1981)
\url{https://doi.org/10.1103/PhysRevD.23.347}

\bibitem{price}
Huw Price,
{\it On the Origins of the Arrow of Time: Why There is Still a Puzzle about the Low Entropy Past},
in Christopher Hitchcock, ed.,
{\it Contemporary Debates in the Philosophy of Science}
(Blackwell, Oxford, 2004)

\bibitem{bek}
Jacob D. Bekenstein,
{\it Black Holes and Entropy},
Phys. Rev. D {\bf 7}, 2333 (1973)
\url{https://doi.org/10.1103/PhysRevD.7.2333}

\bibitem{hawk}
S. W. Hawking,
{\it Black holes and thermodynamics},
Phys. Rev. D {\bf 13}, 191--197 (1976)
\url{https://doi.org/10.1103/PhysRevD.13.191}

\end{thebibliography}
\end{document}